\documentclass[aps,floats,amssymb,twocolumn,prb, superscriptaddress]{revtex4-2}
\usepackage{calc}
\usepackage{psfrag}
\usepackage{graphicx}
\usepackage{color}
\usepackage{bm}
\usepackage{float}
\usepackage{lipsum}
\usepackage{amsmath}
\usepackage[unicode,linktocpage=true]{hyperref}
\hypersetup{colorlinks=true,linkcolor=blue,urlcolor =blue,citecolor=blue,anchorcolor=blue} 
\def\nn{\nonumber}
\def\bs{\boldsymbol}

\begin{document}

\title{Quantum Hall bilayer in dipole representation}

\author{S. Predin}
\affiliation{Scientific Computing Laboratory, Center for the Study of Complex Systems,Institute of Physics Belgrade, University of Belgrade, Pregrevica 118, 11080 Belgrade, Serbia}
\author{M.V. Milovanovi\'c}
\affiliation{Scientific Computing Laboratory, Center for the Study of Complex Systems,Institute of Physics Belgrade, University of Belgrade, Pregrevica 118, 11080 Belgrade, Serbia}

\begin{abstract}
The quantum Hall bilayer (QHB) at filling factor $ \nu = 1 $ represents a competition between Bose-Einstein condensation (BEC) at small distances between layers and fermionic condensation, whose influence grows with distance and results in two separate Fermi liquid states for the underlying quasiparticles at very large (or infinite) distances. The question that can be raised is whether, at intermediate distances between layers, a distinct phase exists, or if a singular transition occurs, with the possibility that this happens at infinite distances. Here, using a dipole representation for fermionic quasiparticles, we find support for the latter scenario: Within a large and relevant range of distances, BEC condensation, identified as Cooper 
$ s $-wave pairing of dipole quasiparticles, prevails over both Cooper 
$ p $-wave pairing and $ s $-wave excitonic pairing of the same quasiparticles.
\end{abstract}

\maketitle

\section{Introduction} 
\label{intro}
The QHB \cite{eimd, ei}  is a fractional quantum Hall effect (FQHE) system with an additional - layer degree of freedom: Two layers of two-dimensional electron gases at distance $d$ from each other, are pierced by a strong magnetic field, $B$, perpendicular to the layers.  The total density of the system, $n_T$, matches the density of available states in the (lowest) Landau level (LL), $(e B)/h$, in which electrons live. 
Thus the total filling factor is $\nu_T = n_T (2 \pi l_B^2) = 1$, with the characteristic length of the system, $l_B = \sqrt{\hbar /(c B)}$, the magnetic length.

Therefore, each layer is half-filled, i.e. it represents a system of electrons that occupy half of the available states in the lowest LL (LLL), and thus $\nu_T = \nu_\uparrow + \nu_\downarrow $,  $\nu_\sigma = 1/2$, $\sigma = \uparrow, \downarrow$, where up, down sign refers to a 
specific layer.  At long-distances, $ d \gg l_B$,  the layers are almost independent, at least 
much less intertwined, and  at $ d \lesssim l_B $, they are strongly coupled and an  excitonic 
binding between a hole in one layer and a particle in the opposite one dominates. Thus,  a 
distance,  ${\bar d} \sim l_B $,  may represent a characteristic distance  for the transition 
from a strong-coupling to a weak-coupling regime, for the system of two layers,  and a question 
can be raised: What happens at these intermediate, $d \sim l_B $ distances? A new intermediate phase, a single transition between two phases (connected with two extremes, small and large distances), or a crossover with no phase transition? Various scenarios appeared in the literature and in this work we will address this question using a special formalism.

The first proposals for multicomponent FQHE systems and studies of the QHB in Ref. \cite{t1,t2,t3,t4}, were followed by experiments \cite{e1,e2} , which confirmed the integer QHE for small distances between the layers in the case of the QHB.  Further development of the theoretical understanding of the system at $d \lesssim l_B$, as an excitonic condensate or an ordered state of the pseudospin of electrons \cite{T1,T2,T3,T4} was followed by the experiments \cite{E1,E2} that revealed the new ordered state and phase at small distances. On the other hand, there is an expectation that at large distances ($d \gg l_B$), we have well separated layers, each in a compresssible state \cite{hlr} of a single layer at filling factor 1/2 \cite{MD,kel2}.

Many theoretical, analytical and numerical, studies have been done \cite{T1,T2,T3,T4,T5,T6,T7,T8,T9,TT0,TT1,TT2,TT3,TT4,TT5,TT6,TT7,TT8,TT9,TTT1,TTT2,TTT3,TTT4,TTT5,TTT6,TTT7,TTT8,TTT9,m1,m2,m3,al,fd,bon1,bon2,sod,hs,gc}, in order to understand the evolution of the QHB with distance, in particular, with the assumption of the projection of the physics into the LLL, in the absence of disorder. The modeling and understanding of the FQHE is based on the composite excitations - particles which are often identified as composite fermions (CFs) \cite{jainp,jainb,oh}.

In the case of the single-layer at filling factor 1/2, and under assumption that all electrons are in the LLL, composite fermion  can be viewed as a composite of electron and its correlation hole, which represents a unit of a positive charge. Thus CF is an overall neutral fermionic object which can make a Fermi sea of CFs (and we may expect a compressible behavior of the system). If we apply a classical analogy, such an overall neutral composite, in an external magnetic field, $B$, must have its momentum proportional to its dipole moment, and at and near Fermi surface we have dipoles \cite{scre1,scre2,rere}. This dipole picture is a direct consequence of the projection into a LL \cite{scre1,scre2,rere}.

Nevertheless, there is an additional feature of the system of electrons that fill half of available states in a LL: The physics should be invariant under exchange of particles and holes i.e. we have a PH (particle-hole) symmetry and, together with CFs, we should also consider and incorporate composite holes (CHs) in our description to have the PH symmetry manifestly represented. To include the PH symmetry, a two component Dirac-type description was introduced in Ref. \cite{son} for the description of half-filled LL.  
But, to describe, in a Fermi-liquid (FL) framework, a half-filled LL of electrons, we may also consider a variant of the dipole construction in a one-component fermion formalism, as introduced in Ref. \cite{pkm}. This construction is a generalization of the dipole i.e. CF representation in the case of bosons at filling factor $\nu = 1$ in the LLL\cite{ph,read}, and it is applicable in the low-energy limit (of the effective FL description). 

In this work, we applied the variant of the dipole representation that we developed \cite{pkm}, in order to understand the evolution of the QHB with distance. We identified, at all distances between layers, the presence of a single phase that can be described as an $s$-wave  Cooper pairing of dipole quasiparticles. The underlying physics of this pairing is the excitonic attraction between electrons in one layer and holes in the opposite layer, and thus the phase that is well-understood at small distances, continues to exist at large distances. 
This scenario was proposed in Ref. \cite{sod}, on the basis of the Dirac description \cite{son} of the physics in each layer, and was numerically supported in a recent work \cite{hs} by modelling the system on a sphere.  This work modeled the underlying physics as an attraction (Cooper pairing) between CF in one layer and CH in the other, opposite layer.
In our study, using the formalism of Ref. \cite{pkm}, the nature of the dominant phase is further elucidated, as is the competition among other candidates for the ground state of the system, as the distance $d$ is varied. 
The $s$-wave Cooper pairing of dipoles prevails over a Cooper $p$-wave pairing and  $s$-wave excitonic pairing of dipole quasiparticles. 
The excitonic pairing  would represent a topological, incompressible phase inside a LLL. Our formalism and the BCS treatment of its set-up are more accurate in the weak-coupling regime, and the confirmation at intermediate and large distances of the same phase (that is dominant in the strong-coupling regime at small distances) is, in this sense, reliable and supports the extension to all $d$'s.

In the dipole formalism that we applied, in the case of bilayer, the dipole quasiparticles 
(which are neither CFs nor composite holes (CHs), but are symmetric objects 
that are consistent with the PH symmetry) enable a 
representation that has manifest symmetry under the exchange of particles and 
holes is done simultaneously in both layers. The dipole representation has additional, 
artificial degrees of freedom - correlation holes, which are identified with holes 
in the electron system(s). This unusual constraint, which we have to 
incorporate into the description, comes from the projection into a single LL of the states of 
fermionic quasiparticles that reside near the Fermi level (and most significantly influence the physics). To incorporate the constraint and use the 
mean-field method, we need to deal with effective Hamiltonians which are adapted to the use of 
the method by explicit inclusion of the constraint (as null operators) in their description. We 
focus (narrow possibilities) on small number of effective Hamiltonians which explicitly 
represent, in their forms, physics of potential phases. We solve them (in the 
mean-field approximation) and compare the energies of different  Hamiltonians 
to find the most stable solution at distance $d$.

The paper is organized as follows. Section \ref{dipole} provides a review of the dipole representation in a single layer.  We discuss the key concepts and principles underlying the dipole representation and its relevance to our study. In Section \ref{bilayer}, we  investigate the implications of the dipole representation in the QHB case, and present results on the competition among phases and the resulting phase diagram. Finally, in Section \ref{conclusion}, we summarize our findings and provide concluding remarks.

\section{Dipole representation for half-filled LL}
\label{dipole} 
The dipole representation for half-filled LL is an extension of the formalism introduced for the description of the CF quasiparticles for a system of bosons at a filling factor $ \nu = 1$ in an isolated LL \cite{ph, read}. In an enlarged space the CF annihilation operator, $ c_{m n}$, is introduced as an operator with double indeces, where each index corresponds to a state in a LL, $n, m = 1, 2, \ldots, N_\phi $, and the left ($L$, physical ) index is associated with a state of an elementary boson, and the right ($R$, artificial)  index is associated with the state of the corresponding correlation hole. In the context of FQHE the correlation hole can be defined by a (local) insertion of flux quanta in the system and represents a well-defined object with charge and statistics. In the system of bosons, the many-body hole is fermionic, and the resulting composite object is a fermion i.e. boson + correlation hole = CF. We may introduce the physical and artificial (of additional degrees of freedom) densities, 
 \begin{equation}
\rho_{n n'}^{L} = \sum_{m} c_{m n}^{\dagger} c_{n' m},
\label{lgen}
\end{equation}
and
\begin{equation}
\rho_{m m'}^{R} = \sum_{n} c_{m n}^{\dagger} c_{n m'},
\label{rgen}
\end{equation}
and their forms in the inverse space,
\begin{equation}
\rho_{\bs{q}}^{L} = \int \frac{d\bs{k}}{(2\pi)^2} c_{\bs{k} - \bs{q}}^\dagger c_{\bs{k}} \exp\left(i \frac{\bs{k} \times \bs{q}}{2}\right), \label{bden}
\end{equation}
and
\begin{equation}
\rho_{\bs{q}}^{R} = \int \frac{d\bs{k}}{(2\pi)^2} c_{\bs{k} - \bs{q}}^\dagger c_{\bs{k}} \exp\left(- i \frac{\bs{k} \times \bs{q}}{2}\right), \label{bdenr}
\end{equation}
which have the same form as the projected densities of systems of elementary particles into a single LL; they are non-local and obey the Girvin-MacDonald-Platzmann algebra.
The collapse from the two-particle to a single-particle index $\bs{k}$ is physically enabled by the existence of the well-defined dipole object (CF) which momentum ($\bs{k}$) in the (external) magnetic field is proportional to its dipole moment.

To complete the description of the system we need to impose constraints in order to have as many degrees of freedom as required by the definition of the problem. It is not hard to see (due to the fact that the total number of CFs is equal to the number of bosons, and due to their fermionic statistics) that we need to have  $\rho_{n n}^{R} = 1$ for each $n$, or $\rho_{\bs{q}}^{R} = 0 $ when $ \bs{q} \neq 0$, in inverse space. In the case of the half-filled LL of electrons, details can be found in \cite{pkm}, we may formally proceed with the same constructions as in the previous case, but now the correlation holes have bosonic statistics. To have a well-defined description we need to impose
$\rho_{n n}^{L} + \rho_{n n}^{R}= 1$ for each $n$, or $\rho_{\bs{q}}^{L} + \rho_{\bs{q}}^{R} = 0 $ when $ \bs{q} \neq 0$ (which requires  that correlation holes are hard-core bosons). But these constraints include the densities of the physical sector and thus have non-trivial influence on the physical degrees of freedom: the correlation holes are on the positions of the real (fermionic) holes. This is an unexpected constraint that opposes the usual interpretation of the correlation hole as a potential well for an elementary particle. Nevertheless, the constraint corresponds to the physics of the CFs near the Fermi level, i.e. to the most important - effective physics of the problem: the magnitude of the momentum of these CFs is $|\bs{k}| \sim k_F = 1/l_B $ ($l_B$ is the magnetic length), and due to the projection to a fixed LL \cite{scre1,scre2,rere}, this implies that the correlation hole is shifted, distanced from 
the electron for the same amount, $|\bs{k}| l_B^2 \sim l_B $. 
Furthermore, we consider the Hamiltonian of the problem, in a PH symmetric form, one that is symmetric under exchange of particles and holes i.e. $L$ and $R$ densities,

\begin{eqnarray}
&& H =  \nn \\
&&\frac{1}{2}\int \frac{d{\bs{q}}}{(2\pi)^2}{\tilde V} (|{\bs{q}}|)   \label{eHam}
\frac{ (\rho^{L}({-\bs{q}}) - \rho^{R}({-\bs{q}}))}{2}
  \frac{(\rho^{L}({\bs{q}}) - \rho^{R}({\bs{q}}))}{2}. \nn \\
\end{eqnarray}
Because of the constraint and the implied PH symmetry, we may refer to the composite object, not as a CF, but simply as a dipole, i.e., a symmetric object which is neither CF nor CH.

\section{The quantum Hall bilayer in dipole representation} 
\label{bilayer}
We begin with the Hamiltonian for the QHB in the second quantization, with electron density operators, $ \rho_{\sigma} ({\bs{q}})$, $ \sigma = \uparrow, \downarrow$ ($\uparrow$ and $\downarrow$ refer to the two different layers) :
\begin{eqnarray}
{\cal H}_e &=& \int \frac{d{\bs{q}}}{(2\pi)^2}  \; \{ \sum_\sigma \frac{1}{2} V(|{\bs{q}}|)  \; \label{2Ham}
: \rho_{\sigma} ({\bs{q}}) \;
 \rho_\sigma (-{\bs{q}}): +  \nn \\
&& V_{\uparrow \downarrow}(|{\bs{q}}|) \;  \label{2Ham}
 \rho_{\uparrow} ({\bs{q}})\;
 \rho_\downarrow (-{\bs{q}}) \}.
\end{eqnarray}
In the enlarged space formalism, the bilinears $ \rho_{\sigma} ({\bs{q}})$ become
\begin{equation}
\rho_\sigma^{L}  ({\bs{q}}) = \int \frac{d\bs{k}}{(2\pi)^2} c_\sigma^\dagger ({\bs{k} - \bs{q}}) c_\sigma ({\bs{k}}) \exp\left(i \frac{\bs{k} \times \bs{q}}{2}\right), \label{bden}
\end{equation}
where formally we have, instead of electron annihilation and creation operators, the quasiparticle operators, $c_\sigma ({\bs{k}}) $ and $c_\sigma^\dagger ({\bs{k}}) $.
Quasiparticles  in the long-distance approximation can be interpreted as fermionic dipoles. 
In the Hamiltonian we recognize the intra-interaction terms with $ V(|{\bs{q}}|) =\left( 1 / |\bs{q}|\right) \exp\left(- |\bs{q}|^2/2\right) $, and the inter-interaction term with $ V_{\uparrow \downarrow}(|{\bs{q}}|) = V(|{\bs{q}}|)  \exp\left(- d |\bs{q}|\right) $, where $d$ denotes the distance between the layers. 
We then proceed by using the dipole representation, which we find optimal for exploring the influence of the fermionic quasiparticles and physics that grows with distance. This representation  allows the inclusion of the PH symmetry of the system (under exchange of all electrons, irrespective of index, and holes) in a manifestly invariant way in the Hamiltonian. 
We then proceed to utilize the dipole representation, which we find optimal for exploring the influence of the fermionic quasiparticles and the physics that evolves with distance. 

By imposing the constraints,
\begin{equation}
 \rho^{L}_\sigma ({\bs{k}}) + \rho^{R}_\sigma ({\bs{k}}) = 0 \;\;\;\;\;\; \sigma = \uparrow, \downarrow ,
\end{equation}
that define the dipole representation in each layer, we place correlation holes where holes are, and thus the PH exchange is followed by the density exchange:
\begin{equation}
 \rho^{L}_\sigma ({\bs{k}})  \leftrightarrow \rho^{R}_\sigma (- {\bs{k}})   \;\;\;\;\;\; \sigma = \uparrow, \downarrow .
\end{equation}
Therefore, the Hamiltonian can be written (by using the constraints) in an explicitly invariant form under this exchange:
\begin{eqnarray}
&& {\cal H}_0 = \nn \\
&& \int \frac{d{\bs{q}}}{(2\pi)^2}  \{ \sum_\sigma \frac{1}{8} V(|{\bs{q}}|)   \label{22Ham}
 (\rho^{L}_\sigma ({-\bs{q}}) - \rho^{R}_\sigma ({-\bs{q}})) (\rho^{L}_\sigma ({\bs{q}}) - \rho^{R}_\sigma ({\bs{q}}))  \nn \\
&&+  \frac{ V_{\uparrow \downarrow}(|{\bs{q}}|)}{4}  \;  \label{2Ham}
 (\rho^{L}_\uparrow ({-\bs{q}}) - \rho^{R}_\uparrow ({-\bs{q}})) (\rho^{L}_\downarrow ({\bs{q}}) - \rho^{R}_\downarrow ({\bs{q}}))
 \}.
\end{eqnarray}
Note the absence of normal ordering due to the requirement that the constraints commute with the Hamiltonian in the physical space. This induces single particle terms (beside purely interacting) with effective mass $M$ (due to the intralayer interaction) \cite{dose,ms}.
By treating the constraints as null operators (in the physical space), which we can include in the Hamiltonian, we reach forms of the Hamiltonian that are adapted to the mean field approach, as they offer obvious interpretation which phase in the mean-field approach they support.

In the QHB case we can add and subtract (product of) constraints and define the following (effective) Hamiltonians:
\begin{eqnarray}
&& {\cal H}_1 = \nn \\
&& \int \frac{d{\bs{q}}}{(2\pi)^2}  \{ \sum_\sigma \frac{1}{8} V(|{\bs{q}}|)   \label{22Ham}
 (\rho^{L}_\sigma ({-\bs{q}}) - \rho^{R}_\sigma ({-\bs{q}})) (\rho^{L}_\sigma ({\bs{q}}) - \rho^{R}_\sigma ({\bs{q}}))  \nn \\
&&+  \frac{ V_{\uparrow \downarrow}(|{\bs{q}}|)}{2}  \;  \label{3Ham}
 (- \rho^{L}_\uparrow ({-\bs{q}}) \rho^{R}_\downarrow ({\bs{q}}) - \rho^{R}_\uparrow ({-\bs{q}}) \rho^{L}_\downarrow ({\bs{q}})  )
 \},
\end{eqnarray}
and
\begin{eqnarray}
&& {\cal H}_2 = \nn \\
&& \int \frac{d{\bs{q}}}{(2\pi)^2}  \{ \sum_\sigma \frac{1}{8} V(|{\bs{q}}|)   \label{4Ham}
 (\rho^{L}_\sigma ({-\bs{q}}) - \rho^{R}_\sigma ({-\bs{q}})) (\rho^{L}_\sigma ({\bs{q}}) - \rho^{R}_\sigma ({\bs{q}}))  \nn \\
&&+  \frac{ V_{\uparrow \downarrow}(|{\bs{q}}|)}{2}  \;  \label{2Ham}
 (\rho^{L}_\uparrow ({-\bs{q}}) \rho^{L}_\downarrow ({\bs{q}})  +  \rho^{R}_\uparrow ({-\bs{q}})  \rho^{R}_\downarrow ({\bs{q}}))
 \}.
\end{eqnarray}
The form of ${\cal H}_1 $ emphasizes the excitonic attraction between densities from opposite layers (electron-hole attraction),which we expect to dominate physics at small distances.
On the level of effective dipoles - composite particles ($c$'s) this will translate to a strong instability to a Cooper pair formation between $c$'s from different layers. On the other hand, the form of ${\cal H}_2 $ is suggestive of excitonic pairing between $c$'s from opposite layers. Indeed, in a mean-field treatment of ${\cal H}_1 $ and $ {\cal H}_2$, these instabilities can be identified as shown in Fig. 1. Details of the mean-field treatment are provided in the Appendix.

\begin{figure}[H]
	\centering
	\includegraphics[scale=.45]{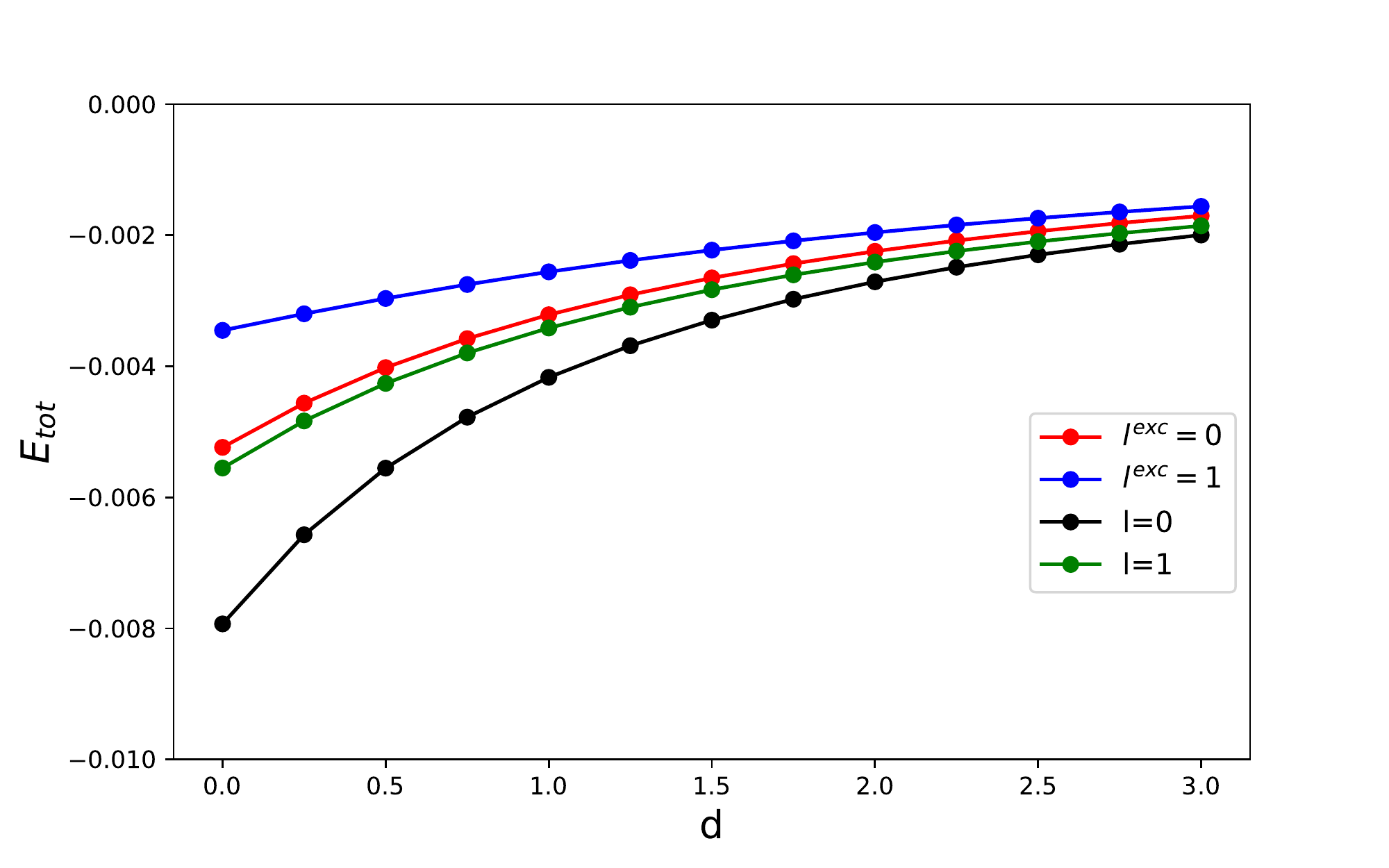}
	\caption{The total energies of the ground states of effective Hamiltonians $H_1$ and $H_2$ as functions of the distance between layers; $s$-wave Cooper pairing of $H_1$ in black,  $p$-wave Cooper pairing of $H_1$ in green, $s$-wave exciton pairing of $H_2$ in red, and $p$-wave exciton pairing of $H_2$ in blue.}
	\label{figure1}
\end{figure}
For all distances considered, the $s$-wave $(l = 0)$ Cooper pairing between layers has lower  energy than the $p$-wave $(l = 1)$ Cooper pairing and $s$-wave $(l = 0)$ excitonic pairing. 
The $ s $-wave excitonic phase of quasiparticles is similar to the one proposed in \cite{al}, which describes an interlayer correlated CF liquid (ICCFL). However, in our work, we operate within a LL, and the quasiparticles involved are (neutral) dipoles.

In calculating total energies we applied a short-distance ``cut-off" if necessary (if we encountered divergences). Though consistently defined on the whole $\bs{k}$ plane \cite{read}, the enlarged space description must be supplied with a natural ``cut-off" (due to an intrinsic ``lattice constant" $l_B$ for this system) : radius $k = \sqrt{2}/ l_B )$ of a circle in the  $\bs{k}$ space i.e. a volume of the available states in a single LL. In most cases  the presence of Gaussians allow the extension of the integration over the whole space.

The exciton instability in the plotted range $ d \in [0, 3 l_B ] $, can be described as an occupation of a single - lower band that is associated with the symmetric superposition: $ ( c_\uparrow (\bs{k}) + c_\downarrow (\bs{k}))/\sqrt{2} $. Thus a single, large Fermi sea exists in this range according to the mean-field calculation. Related to this is the excitonic binding of $c$'s implied by ${\cal H}_0$ (with a dipole-dipole interaction that screens the bare Coulomb interaction) with the gap parameter, $ \Delta_{\bs{k}} \sim |\bs{k}|^2 $ (not a constant as in the case of ${\cal H}_2 $). The total energy of this solution is negligible and may be relevant only for very large $d$, when a transition to two decoupled Fermi seas takes place i.e. for an equal population of symmetric and antisymmetric bands.

We applied the mean-field approach to the effective Hamiltonians and thus we may expect that our results are more reliable for larger $d$ i.e. weak coupling between layers. The weak coupling assumption is completely justified for $d \sim 3 l_B$ (see Fig. 1, weak attraction (pairing amplitudes) i.e. weak coupling can be recognized in small differences in total energies with respect to the asymptotic value i.e. the free-fermion limit), and we may ask whether (at all, because of the strong-coupling for $d \lesssim l_B$) a conclusion for the state of the system at any $d$ can be drawn. But the nature of the predicted phase for $d \lesssim l_B$ (in the strong coupling regime), in our formalism is the same as the one that is firmly confirmed in many calculations and approaches, a binding of the density of electrons with the density of holes in the opposite layers (opposite charge binding), and given that this phase (in our formalism) persists to large $d$ (weak-coupling regime where the approach is fully reliable), we can conclude (assuming continuity i.e.  that a reentrant scenario is unlikely) that one and the same phase is present for all relevant distances including $d \gg l_B$.

In our approach, the binding of charges from opposite layers is described (effectivelly) as $s$-wave binding of dipoles of momenta ${\bs{k}}$ and $-{\bs{k}}$, and thus involves opposite dipole moments from opposite layers. This is similar to the binding described in Ref. \cite{hs}, where a CF in one layer binds to a CH in the other one. The underlying physics of pairing is the same, and the descriptions should correspond to the same phase \cite{comment}.

We did not include a requirement for the boost invariance ($K$-invariance \cite{ms,ss}) as in the single layer case \cite{pkm}, because in this case, a real increase of the energy of the system is possible due to a relative motion between layers. In the mean-field, the dispersion of the Goldstone mode is $\omega_{\bs{k}} \sim \sqrt{\Delta_s / M}$ , where $\Delta_s$ is the BCS gap and $M$ is the mass of quasiparticles due to the Coulomb interaction between the same layer particles. The boost invariance should be ensured in the limiting cases: $ d = 0$ and $d = \infty $, but at intermediate distances, we may rely on the mean-field estimates.

\section{Discussion and conclusions} 
\label{conclusion}
Thus, we may conclude that within the scope of dipole representation and mean-field method, there is no transition in the QHB at finite distance between layers; the excitonic phase of electrons (or Cooper $s$-wave pairing of fermionic quasiparticles - dipoles in the long-distance approximation) dominates the physics at relevant distances.
On the other hand, the suppressed yet competing, $s$-wave excitonic phase of quasiparticle inside LLL  can be described as a large Fermi sea of quasiparticles (with no layer index i.e. with symmetric superpositions).  
An intermediate state in the QHB has been identified and described by exact diagonalization on a torus in \cite{fd}.  
In our description the Fermi sea of (neutral) dipoles, in the absence of the boost invariance (or $K$-invariance), leads to the incompressibility in the charge channel \cite{ms,ss} of the competing phase, and that is consistent with findings of Ref. \cite{fd} (and distinct from the ICCFL phase of Ref. \cite{al}). The exciton condensation induces a gapped, topological behavior in the neutral channel, just as in the case of the ICCFL phase (of Ref. \cite{al}) and consistent with findings of Ref. \cite{fd} . 

Based on previous analyses \cite{bon1,bon2}, one may expect that the inclusion of LL mixing will lead to the formation of the $p$-wave pairing state of CFs (from opposite layers i.e. interpairing). This state can be found in an unprojected (to a fixed LL) Chern-Simons field-theoretical description. A Dirac type of the gauge theory leads to the conclusion that the $p$-wave Cooper pairing of CFs from opposite layers describes the system at any distance between  layers \cite{sod} (except at $d = \infty $). The no transition scenario continues to exist  under LL mixing \cite{gc}.  The results of our work, within the LLL, and the results of Ref. \cite{gc}, with no projection in a Chern-Simons treatment, are in 
correspondence. This continuous (one and only for all distances) phase can be simulated by selecting an appropriate shift (i.e. bias) in the spherical geometry, inside the LLL, as described in \cite{hs}. 

In short, we have demonstrated the usefulness of the dipole representation in the case of the long-standing problem of the QHB, with potential to be used for half-filled problems of general Chern bands. Inside a LLL, the QHB physics is dominated by a single phase: the Cooper $s$-wave pairing of effective dipoles. 
While for the system of electrons in a single layer that occupy half of the available states in a LL  a compressible behavior is only possible \cite{hal}, here we find that a nontrivial double, a nontrivial superposition of two such systems with incompressible, topological behavior based on the noncommutative nature of projection(s) inside LL(s) is in a competition with a compressible phase, but again the compressible  (in the neutral channel) phase is realised. \\

\section*{ACKNOWLEDGMENTS}
We thank Antun Bala{\v z} for discussions on related problems. Computations
were performed on the PARADOX supercomputing facility (Scientific Computing Laboratory, Center for the Study of Complex Systems, Institute of Physics Belgrade). We acknowledge funding provided by the Institute of Physics Belgrade, through the grant by the Ministry of Science, Technological Development, and Innovations of the Republic of Serbia. Furthermore, S.P. is grateful for the financial support received as a returning expert from the Deutsche Gesellschaft für Internationale Zusammenarbeit (GIZ), on behalf of the German Federal Ministry for Economic Cooperation and Development (BMZ).

\section*{appendix}

\section*{Mean-field approach to effective Hamiltonians}

In this Appendix,  we give  a brief account  of the mean-field treatment of the Hamiltonians, ${\cal H}_1$,  (\ref{22Ham}), and ${\cal H}_2$,  (\ref{4Ham}). Due to the attractive nature of the effective interaction in ${\cal H}_1$, and the repulsive  nature of the effective interaction in ${\cal H}_2$, we introduce a mean-field reduction of the Hamiltonians, assuming the pairing in the Cooper channel of ${\cal H}_1$, i.e. $ \langle c_{\bs{k}, \uparrow}^\dagger c_{-\bs{k}, \downarrow}^\dagger \rangle = \Delta_{\bs{k}}^{bcs} \neq 0$, and the excitonic pairing of ${\cal H}_2$, i.e. $ \langle c_{\bs{k}, \uparrow}^\dagger c_{\bs{k}, \downarrow}\rangle = \Delta_{\bs{k}}^{exc} \neq 0$.

In the BCS case, we need to solve self-consistently the following equation, for the order parameter $\Delta_{\bs{q}}^{bcs}$, 
\begin{equation}
\Delta_{\bs{k}}^{bcs} = \int \frac{d{\bs{q}}}{(2\pi)^2}  V_{\uparrow \downarrow}(|{\bs{q}}- {\bs{k}}|) \frac{\Delta_{\bs{q}}^{bcs}}{2 E_{\bs{q}}},
\end{equation}
where $ E_{\bs{q}} = \sqrt{ \xi^2_ {\bs{q}} + |\Delta_{\bs{q}}^{bcs}|^2}$ and $ \xi_ {\bs{q}}  = \epsilon_ {\bs{q}}  -  \epsilon_ {\bs{q}_F}$, with $\bs{q}_F = 1$ (a half-filled condition for each layer). Also  $\epsilon_ {\bs{q}} $ represents the single-particle energy of quasiparticles in each layer that is calculated using the Hartree-Fock method, applied to the single-layer part of the Hamiltonian that describes the intra-layer interaction. See below an explicit formula in (\ref{eps}). The (total) ground state energy of the system is 
\begin{eqnarray}
&&E_0^{bcs}= \nn \\
&&\int \frac{d{\bs{q}}}{(2\pi)^2}   ( \xi_ {\bs{q}} -  E_{\bs{q}})  + \int \frac{d{\bs{q}}}{(2\pi)^2}  \frac{|\Delta_{\bs{q}}^{bcs}|^2}{2 E_{\bs{q}}}.   \nn \\ 
\end{eqnarray}

In the excitonic case, we need to self-consistently solve the following equation, for the order parameter $ \Delta_{\bs{q}}^{exc}$,
\begin{equation}
\Delta_{\bs{k}}^{exc} = \int \frac{d{\bs{q}}}{(2\pi)^2}  V_{\uparrow \downarrow}(|{\bs{q}}- {\bs{k}}|) \frac{\Delta_{\bs{q}}^{exc}}{2 |\Delta_{\bs{q}}^{exc}|} (n_\alpha (\bs{q}) - n_\beta (\bs{q})), \label{SCE}
\end{equation}
where $n_\alpha (\bs{q})$ and $n_\beta (\bs{q})$ denote the occupations of the states with momentum $\bs{q}$, in the band with energies, 
${\cal E}_\alpha (\bs{q}) = \epsilon_ {\bs{q}} - |\Delta_{\bs{q}}^{exc}|$, and in the band with energy, ${\cal E}_\beta (\bs{q}) = \epsilon_ {\bs{q}} + |\Delta_{\bs{q}}^{exc}|$, respectively. In solving (\ref{SCE}), we have to keep the density of the system constant, i.e. the occupation of the lower and upper band, described by appropriate  Fermi momenta, $ q_+^F$ and $ q_-^F$,  should satisfy the following equation, $ (q_+^F )^2 +  (q_-^F)^2 = 2.$ 

The (total) ground state energy of the system is 
\begin{eqnarray}
&&E_0^{exc}= \nn \\
&&\int \frac{d{\bs{q}}}{(2\pi)^2}  [ ( \xi_ {\bs{q}} - |\Delta_{\bs{q}}^{exc}|/2) n_\alpha (\bs{q}) + ( \xi_ {\bs{q}} + |\Delta_{\bs{q}}^{exc}|/2)n_\beta (\bs{q}))], \nn \\ 
\end{eqnarray}
where, as before, $ \xi_ {\bs{q}}  = \epsilon_ {\bs{q}}  -  \epsilon_ {\bs{q}_F}$, with $\bs{q}_F = 1$ (a half-filled condition for each layer). 

The following equation describes the single-particle energy obtained after the application of the (Hartre-)Fock procedure to the intra-layer part of the Hamiltonian:
\begin{eqnarray}
&&\epsilon_ {\bs{k}}= \nn \\
&& \frac{1}{2} \int \frac{d{\bs{q}}}{(2\pi)^2}  {\tilde V} (|{\bs{q}}|)  \left(\sin\left(i \frac{\bs{k} \times \bs{q}}{2}\right)\right)^2 \nn \\
&&-   \int \frac{d{\bs{q}}}{(2\pi)^2}  {\tilde V} (|{\bs{q}}- {\bs{k}}|) \left(\sin\left(i \frac{\bs{k} \times \bs{q}}{2}\right)\right)^2   n (\bs{q}).  \nn \\  \label{eps}
\end{eqnarray}
The occupation $n (\bs{q})$ describes the filled Fermi sphere with radius $q_F = 1$. The first contribution comes from the normal ordering of the density-density form of the intra-layer term, and may represent the self-energy of a dipole \cite{dose}, and the second term represents a Fock contribution.




\begin{thebibliography}{99}
\bibitem{eimd} J. P. Eisenstein, A. H. MacDonald,
\textit{Bose-Einstein condensation of excitons in bilayer electron systems},
\href{https://doi.org/10.1038/nature03081}{ Nature {\bf 432}, 691 (2004)}.

\bibitem{ei} J. P. Eisenstein,
\textit{Exciton Condensation in Bilayer Quantum Hall Systems},
\href{https://doi.org/10.1146/annurev-conmatphys-031113-133832}{ Ann. Rev.  Cond. Mat. Phys. {\bf 5}, 159 (2014)}.

\bibitem{t1} B.I. Halperin, 
\textit{Theory of quantized Hall conductance},
\href{https://www.e-periodica.ch/digbib/view?pid=hpa-001%3A1983%3A56%3A%3A3#3}{Helv. Phys. Acta 56, 75 (1983).}

\bibitem{t2} R.B. Laughlin, 
\textit{Anomalous Quantum Hall Effect: An Incompressible Quantum Fluid with Fractionally Charged Excitations},
\href{https://doi.org/10.1103/PhysRevLett.50.1395}{Phys. Rev. Lett. 50, 1395 (1983).}

\bibitem{t3} F.D.M. Haldane and E.H. Rezayi,
\textit{Fractional quantum Hall effect at even denominator filling factors in multi-layer systems},
\href{}{Bull. Am. Phys. Soc. 32, 892 (1987).}

\bibitem{t4} T. Chakraborty and P. Pietilainen, 
\textit{Fractional Quantum Hall Effect at Half-Filled Landau Level in a Multiple-Layer Electron System},
\href{https://doi.org/10.1103/PhysRevLett.59.2784}{Phys. Rev. Lett. 59, 2784 (1987).}

\bibitem{e1} Y.W. Suen, L. Engel, M.B. Santos, M. Shayegan, and D.C. Tsui,
\textit{Observation of a $\nu$=1/2 fractional quantum Hall state in a double-layer electron system,}
\href{https://doi.org/10.1103/PhysRevLett.68.1379}{ Phys. Rev. Lett. 68, 1379
(1992).}

\bibitem{e2} J.P. Eisenstein, G.S. Boebinger, L.N Pfeiffer, K.W. West, and S. He, 
\textit{New fractional quantum Hall state in double-layer two-dimensional electron systems,}
\href{https://doi.org/10.1103/PhysRevLett.68.1383}{Phys. Rev. Lett. 68,
1383 (1992)}

\bibitem{T1} H.A. Fertig, 
\textit{Energy spectrum of a layered system in a strong magnetic field,}
\href{https://doi.org/10.1103/PhysRevB.40.1087}{Phys. Rev. B 40, 1087-1095 (1989).}

\bibitem{T2} X.G. Wen and A. Zee, 
\textit{Neutral superfluid modes and ‘‘magnetic’’ monopoles in multilayered quantum Hall systems,}
\href{https://doi.org/10.1103/PhysRevLett.69.1811}{Phys. Rev. Lett. 69, 1811 (1992).}

\bibitem{T3} Kun Yang et al., 
\textit{Quantum ferromagnetism and phase transitions in double-layer quantum Hall systems,}
\href{https://doi.org/10.1103/PhysRevLett.72.732}{Phys. Rev. Lett. 72, 732 (1994).}

\bibitem{T4} K. Moon et al., 
\textit{Spontaneous interlayer coherence in double-layer quantum Hall systems: Charged vortices and Kosterlitz-Thouless phase transitions,}
\href{https://doi.org/10.1103/PhysRevB.51.5138}{Phys. Rev. B 51, 5138 (1995).}

\bibitem{E1} I.B. Spielman, J.P. Eisenstein, L.N. Pfeiffer, and K.W. West, 
\textit{Resonantly Enhanced Tunneling in a Double Layer Quantum Hall Ferromagnet,}
\href{https://doi.org/10.1103/PhysRevLett.84.5808}{Phys. Rev. Lett. 84, 5808-5811 (2000).}

\bibitem{E2} M. Kellogg, J.P. Eisenstein, L.N. Pfeiffer, and K.W. West, 
\textit{Vanishing Hall Resistance at High Magnetic Field in a Double-Layer Two-Dimensional Electron System,}
\href{https://doi.org/10.1103/PhysRevLett.93.036801}{Phys. Rev. Lett. 93, 036801 (2004).}

\bibitem{hlr} B. I. Halperin, P. A. Lee, and N. Read,
\textit{Theory of the half-filled Landau level},
\href{https://doi.org/10.1103/PhysRevB.47.7312}{ Phys. Rev. B {\bf 47}, 7312
(1993)}.

\bibitem{MD}  A. H. MacDonald, P. M. Platzman, and G. S. Boebinger, 
\textit{Collapse of integer Hall gaps in a double-quantum-well system,}
\href{https://doi.org/10.1103/PhysRevLett.65.775}{Phys. Rev. Lett. 65, 775 (1990).}

\bibitem{kel2} M. Kellogg, J. P. Eisenstein, L. N. Pfeiffer, and K. W. West,
\textit{Bilayer Quantum Hall Systems at $\nu_T = 1$
: Coulomb Drag and the Transition from Weak to Strong Interlayer Coupling,}
\href{https://doi.org/10.1103/PhysRevLett.90.246801}{Phys. Rev. Lett.{\bf 90}, 246801 (2003)}.

\bibitem{T5} N. E. Bonesteel, I. A. McDonald, and C. Nayak,
\textit{Gauge Fields and Pairing in Double-Layer Composite Fermion Metals,}
\href{https://doi.org/10.1103/PhysRevLett.77.3009}{Phys. Rev. Lett. 77, 3009 (1996).}

\bibitem{T6}  H. Isobe and L. Fu, 
\textit{Interlayer Pairing Symmetry of Composite Fermions in Quantum Hall Bilayers}
\href{https://doi.org/10.1103/PhysRevLett.118.166401}{Phys. Rev. Lett. 118, 166401(2017).}

\bibitem{T7} T. Morinari, 
\textit{Composite-fermion pairing in bilayer quantum Hall systems,}
\href{https://doi.org/10.1103/PhysRevB.59.7320}{Phys. Rev. B 59, 7320 (1999).}

\bibitem{T8} Z. F. Ezawa and G. Tsitsishvili, 
\textit{Quantum Hall ferromagnets,}
\href{https://doi.org/10.1088/0034-4885/72/8/086502}{Reports on Progress in Physics 72, 086502 (2009).}

\bibitem{T9} B. Lian and S.-C. Zhang,
\textit{Wave Function and Emergent SU(2) Symmetry in the $\nu_T = 1 $ Quantum Hall Bilayer,}
\href{https://doi.org/10.1103/PhysRevLett.120.077601}{Phys. Rev. Lett. 120, 077601 (2018).}

\bibitem{TT0} Yong Baek Kim, Chetan Nayak, Eugene Demler, N. Read, and S. Das Sarma,
\textit{Bilayer paired quantum Hall states and Coulomb drag},
\href{https://doi.org/10.1103/PhysRevB.63.205315}{Phys. Rev. B {\bf 63}, 205315 (2000)}.

\bibitem{TT1} Y. N. Joglekar and A. H. MacDonald, 
\textit{Microscopic functional integral theory of quantum fluctuations in double-layer quantum Hall ferromagnets},
\href{https://doi.org/10.1103/PhysRevB.64.155315}{Phys. Rev. B 64, 155315 (2001).}

\bibitem{TT2}  Kun Yang, 
\textit{Dipolar Excitons, Spontaneous Phase Coherence, and Superfluid-Insulator Transition in Bilayer Quantum Hall Systems at $ \nu = 1 $},
\href{https://doi.org/10.1103/PhysRevLett.87.056802}{Phys. Rev. Lett. 87, 196802 (2001).}

\bibitem{TT3} Y. N. Joglekar and A. H. MacDonald, 
\textit{Bias-voltage-induced phase transition in bilayer quantum Hall ferromagnets},
\href{https://doi.org/10.1103/PhysRevB.65.235319}{Phys. Rev. B 65, 235319 (2002).}

\bibitem{TT4} R. Cote, L. Brey, and A. H. MacDonald, 
\textit{Broken-symmetry ground states for the two-dimensional electron gas in a double-quantum-well system},
\href{https://doi.org/10.1103/PhysRevB.46.10239}{Phys. Rev. B 46, 10239 (1992).}

\bibitem{TT5} K. Nomura and D. Yoshioka, 
\textit{Evolution of $ \nu = 1 $ bilayer quantum Hall ferromagnet},
\href{https://doi.org/10.1103/PhysRevB.66.153310}{Phys. Rev. B 66, 153310 (2002).}

\bibitem{TT6} J. Schliemann, S. M. Girvin, and A. H. MacDonald, 
\textit{Strong correlation to weak correlation phase transition in bilayer quantum Hall systems},
\href{https://doi.org/10.1103/PhysRevLett.86.1849}{Phys. Rev. Lett. 86, 1849 (2001).}

\bibitem{TT7} N. Shibata and D. Yoshioka, 
\textit{Fractional Quantum Hall Effects in Graphene and Its Bilayer},
\href{https://doi.org/10.1143/JPSJ.78.104708}{Journal of the Physical Society of Japan 75, 043712 (2006).}

\bibitem{TT8} K. Park, 
\textit{Spontaneous pseudospin spiral order in bilayer quantum Hall systems},
\href{https://doi.org/10.1103/PhysRevB.69.045319}{Phys. Rev. B 69, 045319 (2004).}

\bibitem{TT9} K. Park and S. Das Sarma, 
\textit{Coherent tunneling in exciton condensates of bilayer quantum Hall systems},
\href{https://doi.org/10.1103/PhysRevB.74.035338}{Phys. Rev. B 74, 035338 (2006).}

\bibitem{TTT1} G. Moeller, S. H. Simon, and E. H. Rezayi, 
\textit{Trial wave functions for $ \nu = \frac{1}{2} + \frac{1}{2} $ quantum Hall bilayers},
\href{https://doi.org/10.1103/PhysRevB.79.125106}{Phys. Rev. B 79, 125106 (2009).}

\bibitem{TTT2} G. Moeller, S. H. Simon, and E. H. Rezayi, 
\textit{Paired Composite Fermion Phase of Quantum Hall Bilayers at $ \nu = \frac{1}{2} + \frac{1}{2} $},
\href{https://doi.org/10.1103/PhysRevLett.101.176803}{Phys. Rev. Lett. 101, 176803 (2008).}

\bibitem{TTT3} S. H. Simon, E. H. Rezayi, and M. V. Milovanovic, 
\textit{Coexistence of Composite Bosons and Composite Fermions in $ \nu = \frac{1}{2} + \frac{1}{2} $ Quantum Hall Bilayers},
\href{https://doi.org/10.1103/PhysRevLett.91.046803}{Phys. Rev. Lett. 91, 046803 (2003).}

\bibitem{TTT4} M. V. Milovanovic, E. Dobardzic, and Z. Papic,
\textit{Meron deconfinement in the quantum Hall bilayer at intermediate distances},
\href{https://doi.org/10.1103/PhysRevB.92.195311}{Phys. Rev. B 92, 195311 (2015).}

\bibitem{TTT5} J. Ye, 
\textit{Fractional Charges and Quantum Phase Transitions in Imbalanced Bilayer Quantum Hall Systems},
\href{https://doi.org/10.1103/PhysRevLett.97.236803}{Phys. Rev. Lett. 97, 236803 (2006).}

\bibitem{TTT6} J. Ye and L. Jiang, 
\textit{Quantum Phase Transitions in Bilayer Quantum Hall Systems at a Total Filling Factor $\nu_T=1$ },
\href{https://doi.org/10.1103/PhysRevLett.98.236802}{Phys. Rev. Lett. 98, 236802 (2007).}

\bibitem{TTT7} R. L. Doretto, C. Morais Smith, and A. O. Caldeira, 
\textit{Finite-momentum condensate of magnetic excitons in a bilayer quantum Hall system},
\href{https://doi.org/10.1103/PhysRevB.86.035326}{Phys. Rev. B 86, 035326 (2012)}.

\bibitem{TTT8} R. L. Doretto, A. O. Caldeira, and C. M. Smith,
\textit{Bosonization Approach for Bilayer Quantum Hall Systems at $\nu_T=1$},
\href{https://doi.org/10.1103/PhysRevLett.97.186401}{Phys. Rev. Lett. 97, 186401 (2006).}

\bibitem{TTT9}  B. I. Halperin, in Fractional Quantum Hall Effects: New Developments, edited by B. I. Halperin and J. K. Jain (World Scientific, 2020) pp. 79–132.

\bibitem{m1} Z. Papic, M. V. Milovanovic, Phys. Rev. B {\bf 75}, 195304 (2007).
\textit{Quantum disordering of the 111 state and the compressible-incompressible transition in quantum Hall bilayer systems,}
\href{https://doi.org/10.1103/PhysRevB.75.195304}{Phys. Rev. B {\bf 75}, 195304 (2007)}.

\bibitem{m2} M. V. Milovanovic, Z. Papic, Phys. Rev. B {\bf 79}, 115319 (2009).
\textit{Nonperturbative approach to the quantum Hall bilayer,}
\href{https://doi.org/10.1103/PhysRevB.79.115319}{Phys. Rev. B {\bf 79}, 115319 (2009)}.

\bibitem{m3} M.V. Milovanovic, 
\textit{Paired states in half-filled Landau levels,}
\href{https://doi.org/10.1103/PhysRevB.95.235304}{Phys. Rev. B {\bf 95}, 235304 (2017)}.

\bibitem{al}  J. Alicea, O. I. Motrunich, G. Refael, and M. P. A. Fisher, 
\textit{Interlayer Coherent Composite Fermi Liquid Phase in Quantum Hall Bilayers},
\href{https://doi.org/10.1103/PhysRevLett.103.256403}{Phys. Rev. Lett. {\bf 103}, 256403 (2009)}.

\bibitem{fd} Y. Zhu, L. Fu, and D. N. Sheng, 
\textit{Numerical Study of Quantum Hall Bilayers at Total Filling 
$\nu_{\rm T} = 1 $: A New Phase at Intermediate Layer Distances},
\href{https://doi.org/10.1103/PhysRevLett.119.177601}{Phys. Rev. Lett. {\bf 119}, 177601 (2017)}.

\bibitem{bon1} N. E. Bonesteel,
\textit{Compressible phase of a double-layer electron system with total Landau-level filling factor 1/2},
\href{https://doi.org/10.1103/PhysRevB.48.11484}{Phys. Rev. B {\bf 48}, 11484(R) (1993)}.

\bibitem{bon2} R. Cipri and N. E. Bonesteel
\textit{Gauge fluctuations and interlayer coherence in bilayer composite fermion metals},
\href{https://doi.org/10.1103/PhysRevB.89.085109}{Phys. Rev. B {\bf 89}, 085109 (2014)}.


\bibitem{sod} I. Sodemann, I. Kimchi, C. Wang, and T. Senthil,
\textit{Composite fermion duality for half-filled multicomponent Landau levels},
\href{https://doi.org/10.1103/PhysRevB.95.085135}{Phys. Rev. B {\bf 95}, 085135 (2017)}.

\bibitem{hs} G. Wagner, D. X. Nguyen, S. H. Simon, B. I. Halperin, 
\textit{s
-Wave Paired Electron and Hole Composite Fermion Trial State for Quantum Hall Bilayers with  
$\nu$ = 1},
\href{https://doi.org/10.1103/PhysRevLett.127.246803}{Phys. Rev. Lett. {\bf 127}, 246803 (2021)}.

\bibitem{gc} Luca R{\"u}egg, Gaurav Chaudhary, Robert-Jan Slager,
\textit{Pairing of Composite-Electrons and Composite-Holes in  $\nu_T = 1$ Quantum Hall Bilayers},
\href{https://arxiv.org/pdf/2303.10212.pdf}{ arXiv:2303.10212}.

\bibitem{jainp} J. K. Jain,
\textit{Composite-fermion approach for the fractional quantum Hall effect},
J. K. Jain
\href{https://doi.org/10.1103/PhysRevLett.63.199}{Phys. Rev. Lett. 63, 199 (1989).}

\bibitem{jainb} J. K. Jain,
\textit{ Composite Fermions},
\href{
https://doi.org/10.1017/CBO9780511607561}
{ Composite Fermions (Cambridge University Press, 2007).}

\bibitem{oh}  O. Heinonen, ed., 
\textit{Composite Fermions: A Unified
View of the Quantum Hall Regime},
\href{https://doi.org/10.1142/3894}{Composite Fermions: A Unified
View of the Quantum Hall Regime (World Scientific, 1998).}

\bibitem{scre1} N. Read,
\textit{Theory of the half-filled Landau level},
\href{http://dx.doi.org/10.1088/0268-1242/9/11S/002}{Semi. Cond. Sci. Tech. 9, 1859 (1994)}.

\bibitem{scre2} N. Read,
\textit{Recent progress in the theory of composite fermions near even-denominator filling factors},
\href{https://doi.org/10.1016/0039-6028(96)00318-4}{Surf. Sci., 361/362,
7, (1996)}.



\bibitem{rere} E. Rezayi and N. Read,
\textit{Fermi-liquid-like state in a half-filled Landau level},
\href{https://doi.org/10.1103/PhysRevLett.72.900}{Phys. Rev. Lett. {\bf 72}, 900 (1994)}. 

\bibitem{son} D.T. Son,
\textit{Is the Composite Fermion a Dirac Particle?},
\href{https://doi.org/10.1103/PhysRevX.5.031027}{ Phys. Rev. X {\bf 5}, 031027 (2015)}.

\bibitem{pkm} S. Predin, A. Kne{\v z}evi\'c, M. V. Milovanovi\'c,
\textit{Dipole representation of half-filled Landau level},
\href{https://doi.org/10.1103/PhysRevB.107.155132}{ Phys. Rev. B {\bf 107}, 155132 (2023)}.


\bibitem{ph} V. Pasquier and F. D. M. Haldane, 
\textit{A dipole interpretation of the $\nu = 1/2$ state},
\href{https://doi.org/10.1016/S0550-3213(98)00069-8}{ Nucl. Phys. B {\bf 516}, 719 (1998)}.

\bibitem{read} N. Read, 
\textit{Lowest-Landau-level theory of the quantum Hall effect: The Fermi-liquid-like state of bosons at filling factor one},
\href{https://doi.org/10.1103/PhysRevB.58.16262}{ Phys. Rev. B {\bf 58}, 16262 (1998)}.




\bibitem{dose} Z. Dong and T. Senthil, 
\textit{Noncommutative field theory and composite Fermi liquids in some quantum Hall systems},
\href{https://doi.org/10.1103/PhysRevB.102.205126}{Phys. Rev. B {\bf 102}, 205126 (2020)}.

\bibitem{ms} G. Murthy and R. Shankar,
\textit{Hamiltonian theories of the fractional quantum Hall effect},
\href{https://doi.org/10.1103/RevModPhys.75.1101}{Rev. Mod. Phys. {\bf 75}, 1101 (2003)}.


\bibitem{comment} For example, the wave function of the phase in \cite{hs}, has the characteristic topological number - shift equal to one. If we relate CFs and CHs to the components of the Dirac PH symmetric description of the layers, as argued in Ref. \cite{goc} , the $s$-wave pairing of CFs and CHs of Ref.\cite{hs}, on the level of of the Dirac's eigenstates corresponds to the $p$-wave pairing (of opposite chirality with respect to the one given by the external field) of effective, low-energy fermions of two layers \cite{m3}, and thus we expect such a shift.  Similarly, in our picture,  the intra-layer description is related to  an effective Majorana fermion theory \cite{m4} that leads to $p$-wave pairing as in the previous case. (Our dipoles are made of charges with half unit of electron charge, have a Dirac structure built in \cite{chse}, but are one-component and thus Majorana fermions.) Therefore we expect the same shift.


\bibitem{ss}  S. Simon, 
\textit{The Chern-Simons Fermi liquid description of fractional quantum Hall states,  {\rm Chapter in "Composite Fermions: A Unified View of the Quantum Hall Regime"}},
\href{https://doi.org/10.1142/3894}{Composite Fermions, ed. O. Heinonen, World Scientific}.



\bibitem{goc} D. Go\v{c}anin, S. Predin, M. Dimitrijevi\'{c} \'{C}iri\'{c}, V. Radovanovi\'{c}, M. Milovanovi\'{c},
\textit{Microscopic derivation of Dirac composite fermion theory: Aspects of noncommutativity and pairing instabilities},
\href{https://doi.org/10.1103/PhysRevB.104.115150}{  Phys. Rev. B {\bf 104}, 115150 (2021)}.


\bibitem{m4} M. V. Milovanovi\'c, unpublished


\bibitem{chse} C. Wang, T. Sethil,
\textit{Half-filled Landau level, topological insulator surfaces, and three-dimensional quantum spin liquids}
\href{https://doi.org/10.1103/PhysRevB.93.085110}{Phys. Rev. B {\bf 93}, 085110 (2016)}.



\bibitem{hal} F. D. M. Haldane,
\textit{Incompressible Quantum Hall fluids as Electric Quadrupole fluids},
\href{https://arxiv.org/pdf/2302.12472.pdf}{arXiv:2302.12472}.

















\end{thebibliography}
\end{document}